\documentclass[aps,prd,nofootinbib,showpacs,showkeys,superscriptaddress,twocolumn,10pt]{revtex4}
\usepackage{graphicx}
\usepackage{dcolumn}
\usepackage{bm}
\usepackage{amssymb}
\usepackage{latexsym}
\usepackage{booktabs}
\usepackage[colorlinks, linkcolor=blue, citecolor=blue, urlcolor=blue]{hyperref}

\newcommand{\be}{\begin{equation}}
\newcommand{\ee}{\end{equation}}
\newcommand{\bq}{\begin{eqnarray}}
\newcommand{\eq}{\end{eqnarray}}

\begin{document}

\title{Tilt of primordial gravitational wave spectrum in a universe with sterile neutrinos}

\author{Yun-He Li}
\affiliation{Department of Physics, College of Sciences, Northeastern University, Shenyang
110004, China}
\author{Jing-Fei Zhang}
\affiliation{Department of Physics, College of Sciences, Northeastern University, Shenyang
110004, China}
\author{Xin Zhang\footnote{Corresponding author}}
\email{zhangxin@mail.neu.edu.cn} \affiliation{Department of Physics, College of Sciences,
Northeastern University, Shenyang 110004, China} 
\affiliation{Center for High Energy Physics, Peking University, Beijing 100080, China}

\begin{abstract}
In this work, we constrain the spectral index $n_t$ of the primordial gravitational wave power spectrum in a universe with sterile neutrinos by using the Planck temperature data, the WMAP 9-year polarization
data, the baryon acoustic oscillation data, and the BICEP2 data. We call this model the $\Lambda$CDM+$r$+$\nu_s$+$n_t$ model. The additional massive sterile neutrino species can significantly relieve the tension between the Planck and BICEP2 data, and thus can reduce the possible effects of this tension on the fit results of $n_t$. To constrain the parameters of sterile neutrino, we also utilize the Hubble constant direct measurement data, the Planck Sunyaev-Zeldovich cluster counts data, the Planck CMB lensing data, and the cosmic shear data. We find that due to the fact that the BICEP2 data are most sensitive to the multipole $\ell\sim150$ corresponding to $k\sim0.01$ Mpc$^{-1}$, there exists a strong anticorrelation between $n_t$ and $r_{0.002}$ in the BICEP2 data, and this further results in a strongly blue-tilt spectrum. However, a slightly red-tilt tensor power spectrum is also allowed by the BICEP2 data in the region with larger value of $r_{0.002}$. By using the full data sets, we obtain $m_{\nu,{\rm{sterile}}}^{\rm{eff}}=0.48^{+0.11}_{-0.13}$ eV, $N_{\rm{eff}}=3.73^{+0.34}_{-0.37}$, and $n_t=0.96^{+0.48}_{-0.63}$ for the $\Lambda$CDM+$r$+$\nu_s$+$n_t$ model.
\end{abstract}


\maketitle


\section{Introduction}\label{sec1}

Recently, the BICEP2 Collaboration announced the detection of the primordial gravitational waves (PGWs) with the tensor-to-scalar ratio $r=0.20^{+0.07}_{-0.05}$~\cite{bicep2}. However, the cosmic microwave background (CMB) temperature anisotropy power spectrum measured by the Planck Collaboration gave $r<0.11$ at the 95\% confidence level (CL)~\cite{planck}, which is in tension with the BICEP2 result. To resolve this tension, the BICEP2 Collaboration considered the running of the scalar spectral index, however, the resulting negative running of order $10^{-2}$ significantly challenges the usual slow-roll inflation models. In order to avoid this unsatisfactory consequence, more possible mechanisms to relieve this tension need to be explored. 

We proposed that this tension could be resolved with the consideration of sterile neutrinos~\cite{zx14,Zhang:2014nta} (see also Ref.~\cite{WHu14}).
In our model, two additional parameters, i.e., the effective number of relativistic species, $N_{\rm eff}$, and the effective mass of sterile neutrino, $m_{\nu,{\rm sterile}}^{\rm eff}$, are added to the seven-parameter base $\Lambda$CDM+$r$ model. We call this model the $\Lambda$CDM+$r$+$\nu_s$ model. It has been shown that the $\Lambda$CDM+$r$+$\nu_s$ model successfully yields a large $r$ with Planck data and thus relieves the tension between Planck and BICEP2~\cite{zx14,Zhang:2014nta,WHu14}. Besides, actually, this model can also significantly reduce other tensions between Planck and other astrophysical observations, such as the direct measurement of $H_0$, the Sunyaev-Zeldovich (SZ) cluster counts, and the galaxy shear measurement~\cite{zx14,Zhang:2014nta,WHu14}. It thus appears that the $\Lambda$CDM+$r$+$\nu_s$ model has the potential to become a new cosmological concordance model. For other proposals to produce large B modes, see, e.g., Refs.~\cite{Liu:2014mpa,Harigaya:2014qza,Nakayama:2014koa,Brandenberger:2014faa,Contaldi:2014zua,Miranda:2014wga,Gerbino:2014eqa,McDonald:2014kia,Kehagias:2014wza,Lyth:2014yya,Bonvin:2014xia,Lizarraga:2014eaa,Moss:2014cra,Chluba:2014uba,Cai:2014hja,Li:2014qwa}.

In previous work~\cite{zx14,Zhang:2014nta,WHu14}, the consistency relation $n_t=-r/8$ for single-field slow-roll inflation models was assumed.
Thus, we actually indicate a slightly red-tilt ($n_t<0$) spectrum with $n_t=-0.026$ for our best-fit result $r=0.207$.
However, recently, it has been reported that a strongly blue-tilt ($n_t>0$) spectrum is preferred by the combination of BICEP2 and Planck data, if one takes $n_t$ as a free parameter~\cite{Wang:2014kqa,WuFQ,LiH,Hu:2014aua}. 
On the other hand, by using the BICEP2 data only, Ref.~\cite{Cheng:2014bma} obtained a slightly red-tilt spectrum.
In this paper, we will study the observational constraints on $n_t$ by using the BICEP2 and Planck data in the model with sterile neutrinos. 
We call such a model the $\Lambda$CDM+$r$+$\nu_s$+$n_t$ model, where the tilt $n_t$ is taken as a free parameter.
We will show that our constraint result of $n_t$ is consistent with the results reported in Refs.~\cite{Wang:2014kqa,WuFQ,LiH,Hu:2014aua} 
(and is not in conflict with the results in  Refs.~\cite{Cheng:2014bma,Cheng:2014ota}), i.e., 
a strongly blue-tilt spectrum is preferred by data, but a slightly red-tilt spectrum is also allowed by the BICEP2 data.

This paper is organized as follows. In Sec.~\ref{sec2}, we briefly describe the analysis method and observational data. In Sec.~\ref{sec3}, we present the fit results and discuss them in detail. We give conclusion in Sec.~\ref{sec4}.

\section{Analysis method and data}\label{sec2}

We handle the cosmological perturbations and obtain the theoretical scalar and tensor spectra using the public Boltzmann {\tt CAMB} code\footnote{http://camb.info/}. 
To probe the parameter space, we
use the numerical package {\tt CosmoMC}\footnote{http://cosmologist.info/cosmomc/}
that is based on a Markov Chain Monte Carlo (MCMC) technique.
 The free parameter vector in the $\Lambda$CDM+$r$+$\nu_s$+$n_t$ model is:
$\{\omega_b$, $\omega_c$, $100\theta_{\rm MC}$, $\tau$, $n_s$, $\ln (10^{10}A_s)$, $r_{0.002}$, $n_t$, $m_{\nu,{\rm sterile}}^{\rm eff}$, $N_{\rm eff}\}$,
where $\omega_b\equiv \Omega_b h^2$ and $\omega_c\equiv \Omega_c h^2$ denote the present-day baryon and cold dark matter densities, respectively, $\theta_{\rm MC}$ is the approximation to the ratio of sound horizon to angular-diameter distance to last-scattering surface, $\tau$ is the Thomson scattering optical depth due to reionization, and $n_s$ and $A_s$ are the spectral index and amplitude of the
primordial curvature perturbations, respectively. We use the pivot scale $k_0=0.002$ Mpc$^{-1}$.

For the observations, we use the following data sets. Planck+WP: the CMB temperature power spectrum data from Planck~\cite{planck}, in conjunction with the polarization power spectrum data from 9-year WMAP~\cite{wmap9}. BAO: the latest measurement
of the cosmic distance scale from the Data Release 11 (DR11) galaxy sample of
the Baryon Oscillation Spectroscopic Survey (BOSS),
$D_V(0.32)(r_{d,{\rm fid}}/r_d)=(1264\pm 25)$~Mpc and $D_V(0.57)(r_{d,{\rm fid}}/r_d)=(2056\pm 20)$~Mpc, with
$r_{d,{\rm fid}}=149.28$~Mpc~\cite{boss}.
BICEP2: the CMB angular power spectra (TT, TE, EE, and BB) data from BICEP2~\cite{bicep2}.
$H_0$: the direct measurement of the Hubble constant in the Hubble Space Telescope observations, $H_0=(73.8\pm 2.4)~{\rm km}~{\rm s}^{-1}~{\rm Mpc}^{-1}$~\cite{h0}.
SZ: the combination of $\sigma_8$ and $\Omega_m$ given by the counts of rich clusters of galaxies from the sample of Planck thermal Sunyaev-Zeldovich (SZ) clusters,
$\sigma_8(\Omega_m/0.27)^{0.3}=0.78\pm 0.01$~\cite{tsz}.
Lensing: the CMB lensing power spectrum $C_\ell^{\phi\phi}$ from Planck~\cite{cmblensing},
  and also the combination of $\sigma_8$ and $\Omega_m$ given by the cosmic shear data of the weak lensing from the CFHTLenS survey,
$\sigma_8(\Omega_m/0.27)^{0.46}=0.774\pm 0.040$~\cite{wl}.

In our work, we will use different data combinations to constrain the $\Lambda$CDM+$r$+$\nu_s$+$n_t$ model. 
The Planck+WP+BAO combination is used as a basic data combination since in any case there is no tension between Planck and BAO.
We also use the BICEP2 data in conjunction with the Planck+WP+BAO data to constrain $n_t$.
To constrain the sterile neutrino parameters, we add the $H_0$+SZ+Lensing data into our analysis. 
Besides, we also use the BICEP2 data only to constrain $r_{0.002}$ and $n_t$ for comparison.

\section{Results and discussion}\label{sec3}

\begin{figure*}[tbp]
\centering 
\includegraphics[width=8.1cm]{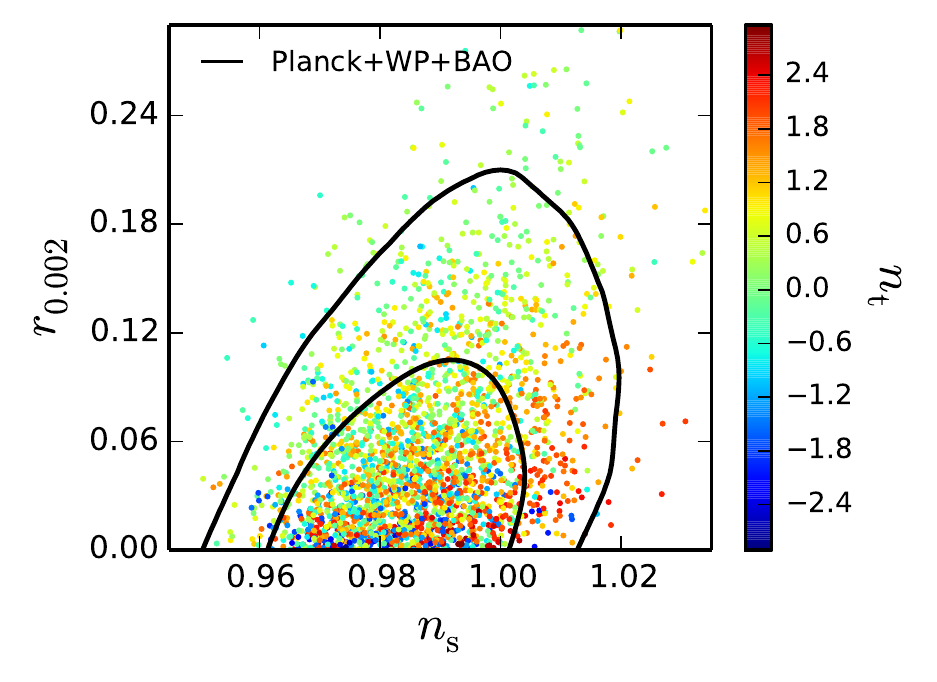}
\includegraphics[width=8cm]{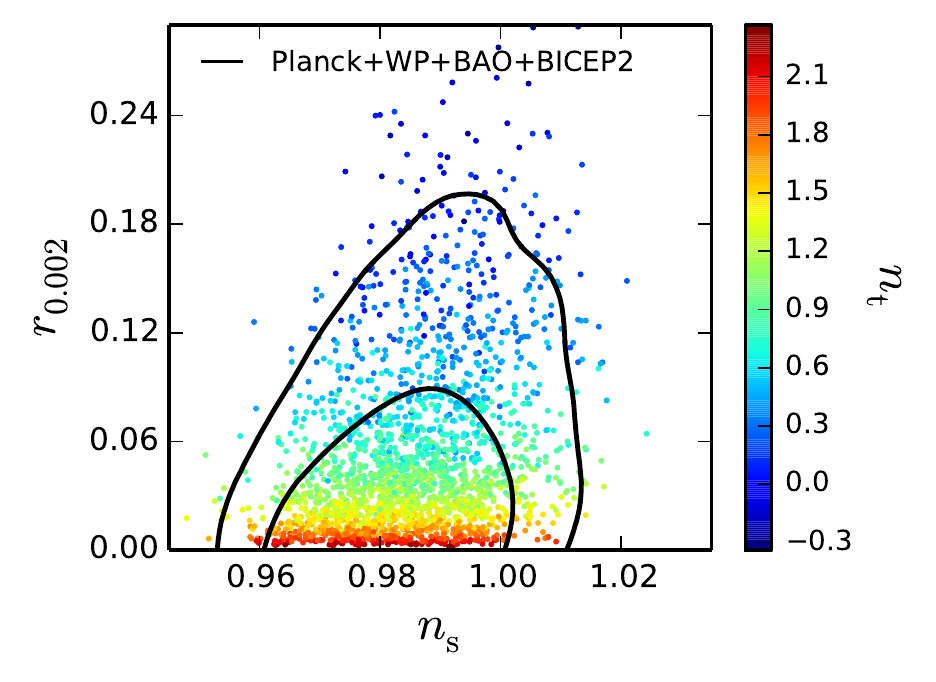}
\hfill
\caption{\label{fig1} Constraint results in the $n_s$--$r_{0.002}$ plane for the $\Lambda$CDM+$r$+$\nu_s$+$n_t$ model by using the Planck+WP+BAO (left panel) and the Planck+WP+BAO+BICEP2 (right panel) data. Points are color-coded according to the values of $n_t$.}
\end{figure*}

Running eight chains with $R-1<0.05$ achieved for each data combinations, we obtain the fit results in the $\Lambda$CDM+$r$+$\nu_s$+$n_t$ model. 

In Fig.~\ref{fig1}, we show the marginalized posterior distribution contours in the $n_s$--$r_{0.002}$ plane by using the Planck+WP+BAO (left) 
and the Planck+WP+BAO+BICEP2 (right) data combinations.
The points are color-coded according to the values of $n_t$. 
From the left panel, one can see that the value of $r_{0.002}$ is amplified in the model with sterile neutrinos by using the Planck+WP+BAO, and 
so the tension between Planck and BICEP2 is alleviated.
Nevertheless, compared to the case with the consistency relation (see Fig.~1 of Ref.~\cite{zx14}), we find that once $n_t$ is free the upper limit of $r_{0.002}$ decreases evidently.
One can therefore find from the right panel that $r=0$ cannot be excluded in the current case (with free $n_t$).
Also, comparing the results of these two panels, we find that the $n_t$ behaves rather differently. 
For example, both the red and blue tilts of PGWs spectrum can be found within the 2$\sigma$ contour in the left panel, while once the BICEP2 data are used (right panel),  
$n_t$ is strongly preferred to be blue and inversely correlates with $r_{0.002}$. 
Other researchers~\cite{Cheng:2014ota,Hu:2014aua} suggest that such a strongly blue-tilt spectrum is the compromise result of the tension between the Planck and BICEP2 data. 
However, we will show that this strongly blue-tilt spectrum is actually induced by the BICEP2 data themselves.

\begin{figure}[tbp]
\centering 
\includegraphics[width=8cm]{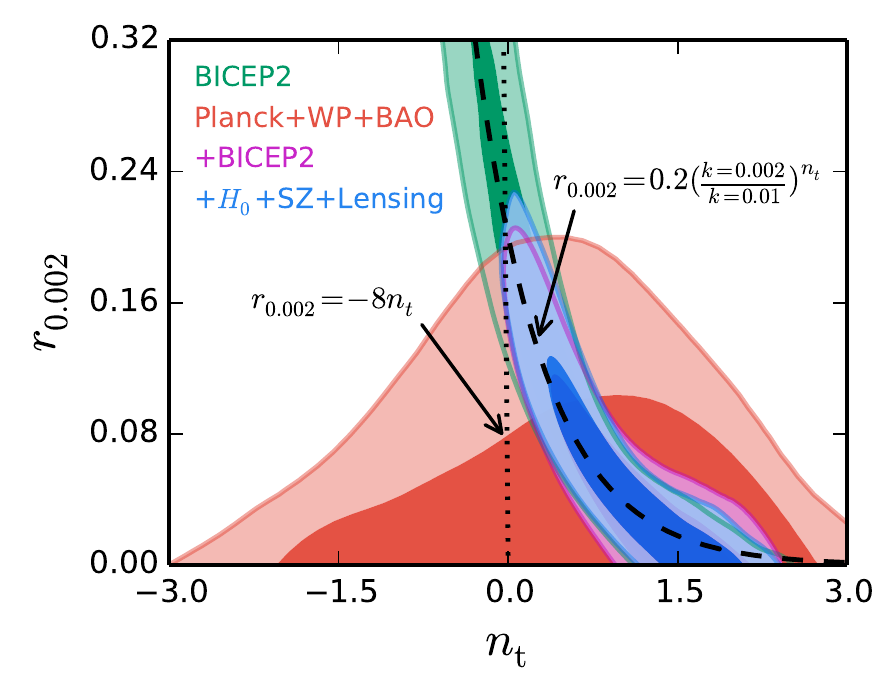}
\hfill
\caption{\label{fig2} Constraint results in the $n_t$--$r_{0.002}$ plane for the $\Lambda$CDM+$r$+$\nu_s$+$n_t$ model by using different data combinations. The black dashed line is plotted for the function $r_{0.002}=0.2(\frac{k=0.002}{k=0.01})^{n_t}$. The black dotted line denotes the consistency relation $r_{0.002}=-8n_t$.}
\end{figure}

In Fig.~\ref{fig2}, we plot the constraint results in the $n_t$--$r_{0.002}$ plane by using different data combinations. We notice that the strong anticorrelation between $n_t$ and $r_{0.002}$ actually comes from the BICEP2 data (see the green contours). This strong anticorrelation between $n_t$ and $r_{0.002}$ in the BICEP2 data can be easily understood. Since the BICEP2 data are most sensitive to multipole $\ell\sim150$ corresponding to $k\sim0.01$ Mpc$^{-1}$, if we fix $r_{0.01}=0.2$, then we immediately obtain $r_{0.002}=0.2(\frac{k=0.002}{k=0.01})^{n_t}$. We show this function by a black dashed line in Fig.~\ref{fig2}. Clearly, this curve is highly consistent with the fit result by using the BICEP2 data only. Now the strong blue-tilt spectrum can be explained as the result that a small value of $r_{0.002}$ cannot be excluded by the BICEP2 data in the case of a free $n_t$. We also test the consistency relation $n_t=-r/8$ (denoted by a black doted line) in Fig.~\ref{fig2}. Using the BICEP2 data only (green contours), we find that a red-tilt spectrum is also preferred in the region with larger $r_{0.002}$. However, once the Planck data are added (magenta and blue contours) in the analysis, the region with a red-tilt spectrum is significantly narrowed. This is due to the fact that the Planck data prefer a small $r_{0.002}$ in the case of free $n_t$.

\begin{table}[tbp]\tiny
\centering\caption{\label{table1} Fit results for the $\Lambda$CDM+$r$+$\nu_s$+$n_t$ model. We quote $\pm 1\sigma$ errors, but
for the parameters that cannot be well constrained, we quote the 95\% CL upper limits.}
\begin{tabular}{lcccccc}
\hline
 &\multicolumn{2}{c}{Planck+WP+BAO+BICEP2} & & \multicolumn{2}{c}{+$H_0$+SZ+Lensing} & \\
\cline{2-3}\cline{5-6}
Parameters & Best fit & 68\% limits && Best fit & 68\% limits & \\
\hline
$\Omega_b$&$0.02229$&$0.02247^{+0.00029}_{-0.00030}$&&$0.02282$&$0.02274\pm0.00029$&\\
$\Omega_c$&$0.1235$&$0.1278^{+0.0054}_{-0.0061}$&&$0.121$&$0.1238^{+0.0052}_{-0.0054}$&\\
$100\theta_{\rm{MC}}$&$1.04074$&$1.04046^{+0.00073}_{-0.00074}$&&$1.0407$&$1.04082^{+0.00077}_{-0.00076}$&\\
$\tau$&$0.096$&$0.098^{+0.013}_{-0.015}$&&$0.108$&$0.107^{+0.015}_{-0.017}$&\\
$m_{\nu,{\rm{sterile}}}^{\rm{eff}}$&$0.00$&$<0.46$&&$0.42$&$0.48^{+0.11}_{-0.13}$&\\
$N_{\rm{eff}}$&$3.45$&$3.73^{+0.32}_{-0.39}$&&$3.56$&$3.73^{+0.34}_{-0.37}$&\\
$n_s$&$0.973$&$0.986\pm0.012$&&$0.989$&$0.990^{+0.014}_{-0.012}$&\\
$n_t$&$1.25$&$1.06^{+0.53}_{-0.64}$&&$1.44$&$0.96^{+0.48}_{-0.63}$&\\
${\rm{ln}}(10^{10}A_s)$&$3.196$&$3.169\pm0.034$&&$3.165$&$3.166^{+0.032}_{-0.031}$&\\
$r_{0.002}$&$0.028$&$<0.154$&&$0.020$&$<0.172$&\\
$\Omega_\Lambda$&$0.6925$&$0.6957^{+0.0094}_{-0.0096}$&&$0.6964$&$0.6945\pm0.0088$&\\
$\Omega_m$&$0.3075$&$0.3043^{+0.0096}_{-0.0094}$&&$0.3036$&$0.3055\pm0.0088$&\\
$\sigma_8$&$0.807$&$0.817^{+0.036}_{-0.030}$&&$0.76$&$0.758\pm0.012$&\\
$H_0$&$69.4$&$70.9^{+1.7}_{-2.1}$&&$70.1$&$70.6^{+1.4}_{-1.8}$&\\

\hline
$-\ln\mathcal{L}_{\rm{max}}$ &\multicolumn{2}{c}{4922.02} & & \multicolumn{2}{c}{4930.77}  & \\
\hline
\end{tabular}
\end{table}

\begin{figure*}[tbp]
\centering 
\includegraphics[width=17cm]{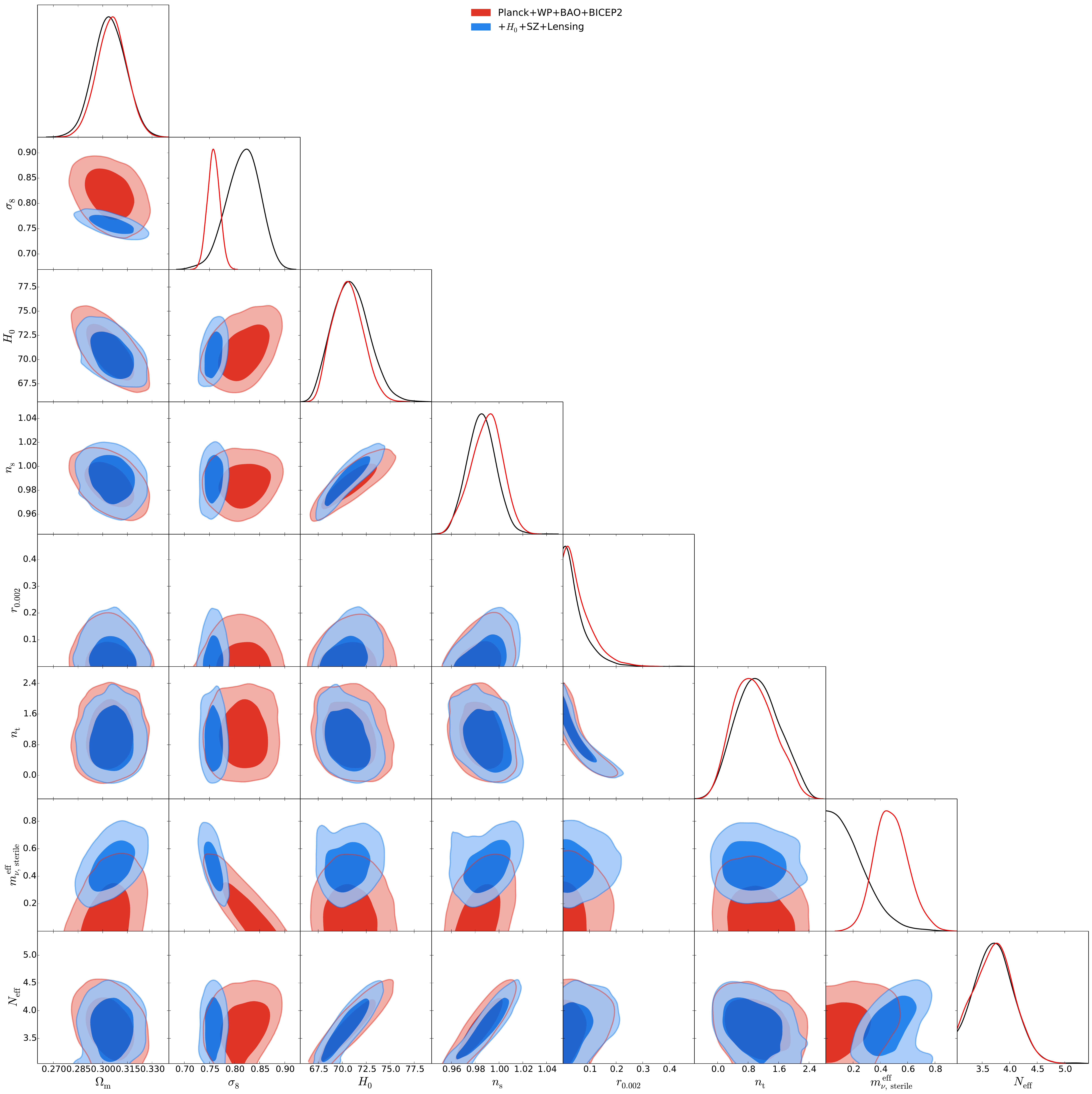}
\hfill
\caption{\label{fig3} Cosmological constraints on the $\Lambda$CDM+$r$+$\nu_s$+$n_t$ model by using the Planck+WP+BAO+BICEP2 data and the Planck+WP+BAO+BICEP2+$H_0$+SZ+Lensing data.}
\end{figure*}

Finally, we summarize all the important fit results for the $\Lambda$CDM+$r$+$\nu_s$+$n_t$ model in Table \ref{table1} and Fig.~\ref{fig3}. By using the Planck+WP+BAO+BICEP2 data combination, we obtain $m_{\nu,{\rm{sterile}}}^{\rm{eff}}<0.46$ eV (95\% CL), $N_{\rm{eff}}=3.73^{+0.32}_{-0.39}$, $r_{0.002}<0.154$ (95\% CL), and $n_t=1.06^{+0.53}_{-0.64}$. The mass of sterile neutrino cannot be well constrained in this case. After adding $H_0$, SZ, and Lensing data, we get tight constraint results for sterile neutrino: $m_{\nu,{\rm{sterile}}}^{\rm{eff}}=0.48^{+0.11}_{-0.13}$ eV and $N_{\rm{eff}}=3.73^{+0.34}_{-0.37}$. We also obtain $r_{0.002}<0.172$ (95\% CL) and $n_t=0.96^{+0.48}_{-0.63}$ by using the full data combination.

\section{Conclusions}\label{sec4}

In this paper, we have studied the observational constraints on the tilt $n_t$ of the PGWs power spectrum in the model with sterile neutrinos by using the Planck+WP, BAO and BICEP2 data.
We call such a model the $\Lambda$CDM+$r$+$\nu_s$+$n_t$ model.
To constrain the parameters of the sterile neutrino species, we also used the $H_0$, SZ, and Lensing data in this work.
We found that there exists a strong anticorrelation between $n_t$ and $r_{0.002}$ in the BICEP2 data.
After a careful analysis, we found that this strong anticorrelation comes from the fact that the BICEP2 data are sensitive to multipole $\ell\sim150$ corresponding to $k\sim0.01$ Mpc$^{-1}$.
Furthermore, due to the anticorrelation between $n_t$ and $r_{0.002}$ in the BICEP2 data, a strong blue-tilt spectrum is preferred, since a small value of $r_{0.002}$ cannot be excluded by the BICEP2 data in the case of a free $n_t$.
We also tested the consistency relation $n_t=-r/8$ in this analysis.
We found that although a strongly blue-tilt spectrum is preferred by the BICEP2 data, a slightly red-tilt spectrum is also allowed in the region with larger value of $r_{0.002}$.
However, since a small $r_{0.002}$ is preferred by the Planck data in the case of free $n_t$, the region with a red-tilt spectrum is significantly narrowed, once the Planck data are added. 
For the $\Lambda$CDM+$r$+$\nu_s$+$n_t$ model, the combination of the full data sets gives $m_{\nu,{\rm{sterile}}}^{\rm{eff}}=0.48^{+0.11}_{-0.13}$ eV, $N_{\rm{eff}}=3.73^{+0.34}_{-0.37}$, $r_{0.002}<0.172$ (95\% CL), and $n_t=0.96^{+0.48}_{-0.63}$.

\begin{acknowledgments}
We acknowledge the use of {\tt CosmoMC}.
This work was supported by the National Natural Science Foundation of
China (Grant No.~11175042) and the National Ministry
of Education of China (Grant No.~N120505003).
\end{acknowledgments}


\end{document}